\documentclass[pdflatex,sn-mathphys-num]{sn-jnl}
\usepackage{graphicx}
\usepackage{amsmath, amssymb}
\usepackage{booktabs}
\usepackage{multirow}
\usepackage{hyperref}
\usepackage{textcomp}

\title{Large Language Models as Implicit Sociological Models:
Reconstructing Voting Behaviour from Sociodemographic Profiles
}

\author*[1,2]{Roman Neruda}\email{roman@cs.cas.cz}
\author[1,2]{Martin Bakoš}\equalcont{These authors contributed equally to this work.}
\author[2]{Josef Šlerka}\equalcont{These authors contributed equally to this work.}
\author[2]{V\'\i t Tuček}\equalcont{These authors contributed equally to this work.}
\author[1,2]{Petra Vidnerová}\equalcont{These authors contributed equally to this work.}
\author[1,2,3]{Gabriela Kadlecová}\equalcont{These authors contributed equally to this work.}

\affil[1]{Institute of Computer Science, Czech Academy of Sciences, Prague, Czech Republic}
\affil[2]{Faculty of Arts, Charles University, Prague, Czech Republic}
\affil[3]{distil labs, Berlin, Germany}

\abstract{
Large language models (LLMs) trained on large-scale internet corpora encode extensive statistical regularities about social identities, attitudes, and political behaviour. This paper introduces and evaluates a methodological framework that leverages these latent representations to reconstruct aggregate voting behaviour from individual-level sociodemographic profiles. We operationalize LLMs as implicit sociological models by conditioning them on demographic descriptions, eliciting probabilistic turnout and party preferences, and aggregating individual outputs via a soft voting procedure. Using the 2021 Czech parliamentary election as a validation case, we demonstrate that contemporary LLMs reproduce official election outcomes with low mean absolute error, recover known political bloc structures, and align with independently established sociodemographic gradients. The contribution of this work is methodological rather than predictive: we show how LLMs can be systematically interrogated as compressed representations of social reality, offering a novel exploratory instrument for computational social science while clearly delineating its epistemic and ethical limits.
}


\begin{document}

\maketitle

\noindent\textbf{Author profile links:}\\
Roman Neruda: Google Scholar: \href{https://scholar.google.cz/citations?user=bZRKRCMAAAAJ}{bZRKRCMAAAAJ}; ORCID:
\href{https://orcid.org/0000-0003-2364-5357}{0000-0003-2364-5357}
\\
Martin Bakoš: Google Scholar: NA; ORCID:
\href{https://orcid.org/0009-0007-4471-7846}{0009-0007-4471-7846}
\\
Josef Šlerka: Google Scholar: \href{https://scholar.google.cz/citations?user=nlHcbnEAAAAJ}{nlHcbnEAAAAJ}; ORCID:
\href{https://orcid.org/0000-0003-3564-1767}{0000-0003-3564-1767}
\\
V\'\i t Tuček: Google Scholar: \href{https://scholar.google.cz/citations?user=vZdgAuAAAAAJ}{vZdgAuAAAAAJ}; ORCID: 
\href{https://orcid.org/0000-0003-0860-6296}{0000-0003-0860-6296}
\\
Petra Vidnerová: Google Scholar: \href{https://scholar.google.cz/citations?user=Leez938AAAAJ}{Leez938AAAAJ}; ORCID: 
\href{https://orcid.org/0000-0003-3879-3459}{0000-0003-3879-3459}
\\
Gabriela Kadlecová: Google Scholar: \href{https://scholar.google.cz/citations?user=KGwHzzMAAAAJ}{KGwHzzMAAAAJ}; ORCID: 
\href{https://orcid.org/0000-0002-4780-0633}{0000-0002-4780-0633}

\section{Introduction}

In the social sciences, political behaviour is among the most thoroughly documented correlates of sociodemographic position. As demonstrated in Lipset and Rokkan's cleavage theory \cite{lipset1967cleavages} and the contemporary documentation of education-based realignment across 21 democracies \cite{gethin2022brahminleft}, decades of empirical research have established that a person's demographic profile substantially shapes their likely vote. These regularities are not straightforward: education and income now predict party support in opposing directions, and centrist populist parties such as ANO in the Czech Republic draw heterogeneous voter bases poorly explained by demographics alone \cite{havlik2018centristpopulism}.

Large language models trained on internet-scale corpora represent a novel form of cultural compression \cite{buttrick2024compression}. Rather than encoding explicit rules or theories, these models internalize statistical regularities in language use, including associations between social identities and political preferences. A mounting body of evidence suggests that this encoding is not merely incidental noise but structured internal representation: word embeddings absorb human-like social biases as a fundamental property of distributional learning \cite{caliskan2017weat}, conversational models exhibit systematic political orientations that vary across model families and can be steered via fine-tuning \cite{rozado2024politicalpreferences}, and mechanistic probing reveals approximately linear ideological dimensions in model activation space that predict both lawmakers' voting records and media slant \cite{kim2025linearpolitics}. The bias literature thus identifies a limitation that also implies a capability: LLMs can be conceptualized as compressed encodings of sociopolitical reality that can, in principle, be systematically interrogated.

This raises an intriguing question: to what extent do LLMs implicitly encode sociopolitical structures well enough to reconstruct aggregate voting behaviour from demographic information alone? In this study, we explore this question using the 2021 Czech parliamentary election as a case study. We condition multiple LLMs on individual-level sociodemographic profiles drawn from a representative survey and ask the models to produce probabilistic voting choices. Aggregating these outputs using a soft voting procedure, we compare the resulting simulated election outcomes with both survey self-reports and official election results.

The Czech case provides a demanding test. Nearly all existing LLM electoral simulations target the US two-party system, where training data is abundant and political structure is relatively simple \cite{argyle2023outofone}. The few non-US studies reveal strong context-dependent failures, European Parliament simulations show uneven accuracy across countries and languages \cite{vonderheyde2024unitedindiversity}, and multilingual evaluations document systematic political-orientation shifts by language, including in Czech \cite{gurgurov2025multilingualpoliticalviews}. Czech is a medium-resource language with documented performance gaps relative to English across standard benchmarks \cite{fajcik2025benczechmark}, and the Czech party system features multiparty competition, post-communist legacies, and high secularism, features that may not be well-captured by models trained predominantly on Anglophone data. If LLMs encode sociopolitical regularities that generalise beyond their training core, this should be detectable even in this challenging context; documenting where they fail is as informative as where they succeed.

This paper makes a methodological contribution to computational social science by proposing and empirically validating a general framework for using large language models as implicit sociological models. Rather than treating LLMs as predictors or explanatory agents, we conceptualize them as high-dimensional compressors of social regularities that can be interrogated via controlled demographic conditioning and probabilistic aggregation (cf. \cite{epstein2008whymodel,buttrick2024compression}). We do not assume individual-level accuracy, indeed, individual LLM predictions are known to be noisy and systematically biased \cite{bisbee2024syntheticreplacements}. Instead, we employ soft voting aggregation over probabilistic outputs, testing whether individual errors cancel sufficiently at the population level to reconstruct aggregate vote shares.

\subsection{Contributions}

This paper makes three primary contributions:

\begin{enumerate}
\item We introduce a general methodological framework for demographic conditioning of large language models to simulate aggregate social behaviour, operationalising the concept of LLMs as implicit sociological models.
\item We formalise a soft voting aggregation procedure that transforms individual-level probabilistic outputs into stable macro-level estimates, designed to handle individual-level noise while remaining diagnostic of systematic biases.
\item 
We provide multi-level validation of the framework, including aggregate vote shares, covariance structure, comparison with polling agencies, and alignment with expert-coded ideological positions (CHES), in a non-Anglophone multiparty context.
\end{enumerate}
\section{Related Work}

\subsection{Sociodemographic voting research}

The relationship between sociodemographic position and party choice is among the best-documented regularities in political science. Lipset and Rokkan's cleavage theory \cite{lipset1967cleavages} identified structural divides, class, religion, centre–periphery, that crystallised into European party systems. This framework has been extended and updated: Gethin, Martínez-Toledano, and Piketty \cite{gethin2022brahminleft} document the emergence of "multi-elite" systems across 21 democracies, where education and income now predict party support in opposing directions. In the Czech context, however, centrist populist parties such as ANO draw heterogeneous voter bases better predicted by political cynicism than by standard sociodemographic variables \cite{havlik2018centristpopulism}, suggesting that demographic conditioning will reconstruct some parties more faithfully than others. We ask whether LLMs, trained on vast corpora of human-generated text, have absorbed these documented regularities with sufficient fidelity to reconstruct aggregate voting outcomes when conditioned on individual sociodemographic profiles.

\subsection{Computational approaches to voting inference}

Understanding how demographic structure shapes electoral outcomes has motivated a range of computational methods. Multilevel regression with poststratification (MRP) and its extensions, including flexible machine-learning variants and deep hierarchical models, represent the dominant statistical approach, but all face a common challenge: the analyst must specify or discover relevant variable interactions, which becomes combinatorially demanding as demographic dimensions multiply \cite{goplerud2024deephierarchicalmrp}. In parallel, text-as-data methods have established language as a key observable for modelling political structure, extracting ideological positions, framing, and sentiment from textual corpora \cite{gentzkow2019textasdata}. Social media platforms have further enabled population-scale studies of political information exposure and polarisation using digital trace data \cite{bakshy2015facebookdiverse}, though such studies are constrained by platform selection bias and limited demographic information. Our approach differs from all three traditions: rather than fitting an explicit model to labelled outcomes, engineering features from text, or relying on platform data, we query a pretrained generative model as an implicit mapping from demographic profiles to probabilistic vote choices.

\subsection{LLMs as social simulators}

The foundational ``silicon sampling" framework of Argyle et al. \cite{argyle2023outofone} conditions GPT-3 on demographic backstories drawn from real survey respondents and demonstrates ``algorithmic fidelity", demographically correlated responses reflecting the interplay of attitudes and sociocultural context. Subsequent work has expanded this paradigm to interactive generative agents with memory and planning \cite{park2023generativeagents} and to election-oriented simulations in both US and European multiparty contexts. At the same time, rigorous critical work warns against overinterpreting aggregate alignment. Bisbee et al. \cite{bisbee2024syntheticreplacements} show that while synthetic survey means can match population averages, LLM responses exhibit substantially compressed variance, and nearly half of regression coefficients diverge significantly from human data, with signs flipping in one-third of cases. Gao et al. \cite{gao2025takecaution} demonstrate more broadly that LLMs fail to replicate human behaviour distributions in strategic settings despite surface-level similarity, cautioning against their use as uncritical human surrogates. These findings establish that success at coarse aggregation does not guarantee inferential validity.

Our work builds on this literature while differing in several respects. We target a non-US multiparty election, a substantially harder setting. We elicit full probability distributions over party choices rather than discrete responses, employing soft voting aggregation designed to handle individual-level noise. We validate not only against aggregate vote shares but also against covariance structure, expert-coded ideological positions, and polling-agency baselines. And we frame the exercise as methodological reconstruction, asking what sociopolitical regularities LLMs have absorbed, rather than claiming that they can replace surveys.

\subsection{Encoded political structure}

A growing body of evidence demonstrates that large language models encode durable social and political regularities present in their training data. This begins with the foundational observation that distributional word representations absorb human-like biases as a fundamental property of learning from language \cite{caliskan2017weat}. The "stochastic parrots" critique identifies the core mechanism, statistical absorption of training-corpus patterns, as a source of risk \cite{bender2021stochasticparrots}, while quantitative evaluations reveal that RLHF-tuned models align most closely with liberal, high-income, well-educated respondents, with persistent misalignment for underrepresented groups even after explicit demographic steering \cite{santurkar2023whoseopinions}. Comparative measurement across 24 models confirms that most conversational LLMs generate left-of-centre responses, though they can be steered to occupy any region of the political spectrum \cite{rozado2024politicalpreferences}. Most directly relevant to our approach, mechanistic probing reveals that political perspective corresponds to approximately linear directions in model activation space, suggesting that ideological dimensions are not merely surface-level text patterns but internally represented geometric structures \cite{kim2025linearpolitics}. For methodological work, these findings can be reinterpreted as evidence that LLMs implement an implicit mapping from context, including demographic signals, to plausible beliefs and choices, a mapping we exploit and validate through electoral reconstruction.

\subsection{Non-English contexts}

Nearly all LLM electoral simulation targets the US two-party system. Recent work that moves beyond this setting reveals that language and context are major sources of bias and failure: a European Parliament study finds that demographic prompting alone performs poorly and unevenly across countries and languages \cite{vonderheyde2024unitedindiversity}, and multilingual ideology measurement across 14 languages, including Czech, shows that political orientation varies significantly by language and model scale \cite{gurgurov2025multilingualpoliticalviews}. Czech-centric benchmarks confirm that Czech competence cannot be inferred from English results, with documented performance gaps across tasks \cite{fajcik2025benczechmark}. Within this landscape, our work contributes a tightly scoped, non-US, multiparty validation case. The Czech Republic, a post-communist society with a fragmented party system and a medium-resource language, represents a genuine out-of-distribution test for models trained predominantly on Anglophone data.

\section{Data}

\subsection{Survey Data}

The primary dataset for this study is derived from the ``Society of Distrust''  survey, a comprehensive study comprising $N=3,880$ respondents, designed to be representative of the Czech adult population. The survey dataset contains answers to more than 130 questions covering a wide range of topics, including sociodemographics, trust in institutions, media consumption habits, and susceptibility to conspiracy narratives. Crucially for this study, it captures political attitudes and self-reported voting behaviour during the 2021 parliamentary election.

\subsection{Validation Targets}

To evaluate the accuracy of our election simulation, we establish ground truth at two distinct levels.

\begin{itemize}
    \item \textbf{Actual Vote:} The official election results published by the Czech Statistical Office, at the regional and district levels.
    \item \textbf{Claimed Vote:} Election results based on self-reported vote choice obtained directly from the survey.
\end{itemize}

This sets the groundwork for a triangular evaluation, allowing a comparative analysis of simulated outcomes, claimed voter behavior, and actual election results.

\subsection{Input Variables}

The simulated elections are conditioned on a feature set of 10 variables, primarily comprising standard sociodemographic indicators in electoral research. These features were selected based on their availability in the dataset and their established theoretical relevance to voting behaviour.  

Two of the ten variables, subjective living standard and interest in politics, are attitudinal rather than structural. Their inclusion is motivated by a well-established finding in the electoral behaviour literature: subjective evaluations of one's economic position consistently outperform objective indicators as individual-level predictors of vote choice, functioning as a politically filtered interpretation of material reality that predicts support for populist parties through mechanisms of relative deprivation (\cite{ferwerda2025nostalgicdeprivation}). Political interest, in turn, strongly structures electoral participation itself (\cite{brady1995beyondses}), while political alienation and cynicism disproportionately drive voters away from established parties and towards new populist or protest movements \cite{havlik2018centristpopulism}. Including these attitudinal covariates allows us to test whether LLMs encode not only structural sociodemographic regularities but also the attitudinal dimensions that mediate between demographic position and partisan preference.

The input vector includes:

\begin{itemize}
    \item \textbf{Demographics:} Gender, Age, Education level.
    \item \textbf{Geography:} Region (\emph{Kraj}), District (\emph{Okres}), Municipality size.
    \item \textbf{Socioeconomics:} Employment status, Household income range.
    \item \textbf{Attitudinal:} Subjective living standard, Interest in politics.
\end{itemize}

While the survey includes extensive data on specific political stances (e.g., attitudes toward NATO or the Russo-Ukrainian war), these were deliberately excluded from the input features to focus on predicting voting behaviour solely from sociodemographic and general background profiles.

\section{Methods}

\subsection{Prompt Design}

To simulate individual voting behaviour, we constructed a prompt for each respondent. This prompt encodes the selected sociodemographic variables into a natural language description of a respondent, concluding with a completion task to estimate the vote choice in the 2021 parliamentary election.

The simulations were conducted entirely in the Czech language to ensure the linguistic context aligns with the socio-cultural setting of the election. Crucially, each simulation was executed as a \textbf{stateless, independent query}. The model retains no memory of previous respondents, ensuring that each prediction is generated in isolation without cross-sample contamination.

An example of a translated respondent prompt is provided below:

\begin{quote}
``I am a man, 43 years old, my education is primary and secondary without graduation. I live in the Central Bohemian region, Beroun district, in a municipality of less than 1,000 inhabitants. I am a full-time employee and our household income is 40,001–60,000 CZK. I have neither good nor bad living standard. I am rather not interested in politics. In the 2021 parliamentary elections, I voted for:''
\end{quote}

We evaluated four OpenAI models to test performance across different capability levels and architecture versions. The specifications for these models, as reported by OpenAI at the time of running the experiments, are detailed in Table~\ref{tab:model_specs}.

\begin{table}[h]
\centering
\caption{Model Specifications}
\label{tab:model_specs}
\begin{tabular}{lccc}
\toprule
\textbf{Model} & \textbf{Context Window} & \textbf{Max Output Tokens} & \textbf{Knowledge Cutoff} \\
\midrule
GPT-4o-mini & 128,000 & 16,384 & Oct 2023 \\
GPT-4o & 128,000 & 16,384 & Oct 2023 \\
GPT-4.1-nano & 1,047,576 & 32,768 & Jun 2024 \\
GPT-4.1 & 1,047,576 & 32,768 & Jun 2024 \\
\bottomrule
\end{tabular}
\end{table}

To ensure reproducibility and minimize stochasticity, all models were queried with the \texttt{temperature} parameter set to 0. All other parameters were left at their default values. The complete configuration, reflecting the default parameter values defined in the OpenAI documentation at that time, is listed in Table \ref{tab:api_params}.

\begin{table}[h]
\centering
\caption{OpenAI API Configuration Parameters}
\label{tab:api_params}
\begin{tabular}{ll}
\toprule
\textbf{Parameter} & \textbf{Value} \\
\midrule
\texttt{temperature} & 0 \\
\texttt{top\_p} & 1 \\
\texttt{parallel\_tool\_calls} & \texttt{true} \\
\texttt{truncation} & \texttt{disabled} \\
\texttt{service\_tier} & \texttt{auto} \\
\texttt{store} & \texttt{true} \\
\texttt{stream} & \texttt{false} \\
\texttt{background} & \texttt{false} \\
\texttt{conversation} & \texttt{null} \\
\texttt{stream\_options} & \texttt{null} \\
\bottomrule
\end{tabular}
\end{table}

\subsection{Structured Output}
A strict output schema was enforced via the API's Structured Outputs feature. Instead of generating unstructured text, the models were required to return a structured JSON object validated against a Pydantic schema. The schema contains two probability distributions:

\begin{itemize}
\item Probability of turnout:
\[
P(\text{voted}), \; P(\text{not voted})
\]
\item Conditional probability distribution over party choices:
\begin{align*}
& P(p \mid \text{voted}), \\
& \text{where} \quad p \in \{\text{ANO}, \text{SPOLU}, \text{PirSTAN}, \text{SPD}, \text{Přísaha}, \text{ČSSD}, \text{KSČM}, \text{Trikolóra}, \text{Other}\}
\end{align*}
\end{itemize}

\subsection{Soft Voting Aggregation}

To aggregate individual predictions into election outcomes, we employ a \textbf{soft voting} strategy. Rather than forcing a discrete choice (hard voting), this approach utilizes the full probability distribution generated by the model, allowing each synthetic respondent to contribute a weighted vote to multiple parties proportional to their predicted likelihood. This preserves the nuances of voter preference and uncertainty inherent in the distribution.

For each respondent $i$ and political party $p$, we define a soft vote weight $w_{i,p}$ as the product of the conditional probability of choosing that party and the predicted probability of turnout:
\[
w_{i,p} = P(p \mid i) \cdot P(\text{voted} \mid i)
\]

The aggregate vote share $\hat{V}_p$ for a specific party $p$ within a target geographic unit (e.g., region or country) is computed by normalizing the sum of these weights against the total weight of all parties $S$:

\begin{equation}
\hat{V}_p = \frac{\sum_{i} w_{i,p}}{\sum_{q \in S} \sum_{i} w_{i,q}}
\end{equation}

\subsection{Evaluation Metrics}

Our primary metric for vote share accuracy is the Mean Absolute Error (MAE). This is calculated by comparing the simulated vote shares ($\hat{V}$) against the ground truth ($V$)—defined as either the official election results or the self-reported survey data.

\begin{equation}
\text{MAE} = \frac{\sum_{p} | \hat{V}_p - V_p |}{N_{parties}}
\end{equation}

Beyond prediction accuracy, we evaluate whether the models recover the latent dependencies of the Czech political landscape, such as the ideological clustering of coalition blocs. To this end, we analyze the covariance and correlation matrices of predicted vote shares, comparing the simulated inter-party relationships against those observed in the official election results.

\subsection{Model Prior Knowledge}\label{sec:model_prior}

Complementary to the sociodemographic simulations, we sought to quantify the models' internal prior knowledge and inherent biases regarding the election results. This baseline assessment allows us to distinguish between the signal derived from the specific persona descriptions and the models' underlying training data distribution.

We evaluated two distinct prompting strategies, adapted for both the national level (Czechia) and individual administrative regions:

\begin{enumerate}
    \item \textbf{Citizen Prior:} A generic persona prompt representing an unspecified resident.
    \begin{quote}
    ``I am a resident of Czechia. In the 2021 parliamentary election, I voted for:''
    \end{quote}
    
    \item \textbf{Direct Prior:} A direct query asking for a factual prediction or retrieval of the election outcome.
    \begin{quote}
    ``The turnout and result of the 2021 parliamentary election in Czechia will be:''
    \end{quote}
\end{enumerate}

To ensure robust baseline estimates, each prompt was executed with $N=10$ repetitions. Although temperature was fixed at 0, prompts were repeated to account for potential residual nondeterminism in model serving and structured decoding. The final prior is calculated as the average of the returned probability distributions across these repetitions.

These baseline queries utilized the same structured output schema (Pydantic) and API parameters as the main simulation to ensure comparability. In the subsequent analysis, these baselines are referred to as the \textbf{Citizen} (generic resident) and \textbf{Direct} (objective prediction) models.

\section{Results}

\subsection{Model Output Characteristics}

Before evaluating predictive accuracy, we first characterize the raw data generated by the simulation. Although all four models successfully returned valid Pydantic objects for every respondent, an initial examination revealed differences in how they handled non-voters. These distributional behaviors are summarized in Table~\ref{tab:output_characteristics}.

\begin{table}[h]
\centering
\caption{Distribution of Absolute Turnout Predictions and Undefined Party Preferences}
\label{tab:output_characteristics}
\begin{tabular}{lccc}
\toprule
\textbf{Model} & \textbf{$P(\text{voted}) = 0$} & \textbf{Undefined Preference} & \textbf{Overlap (Undefined $\cap\ P=0$)} \\
\midrule
GPT-4o-mini & 2.65\% & 2.65\% & 100.0\% \\
GPT-4o & 0.00\% & 0.00\% & -- \\
GPT-4.1-nano & 61.37\% & 62.14\% & 95.0\% \\
GPT-4.1 & 0.00\% & 0.00\% & -- \\
\bottomrule
\end{tabular}
\caption*{Note: ``Undefined Preference'' refers to instances where the model assigned a probability of 0 to all political parties or did not provide any. ``Overlap'' denotes the percentage of undefined preference cases that were simultaneously predicted as absolute non-voters ($P=0$).}
\end{table}

The first notable difference emerged in the assignment of turnout probabilities. The larger models (GPT-4o and GPT-4.1) never assigned an absolute zero probability to voting, maintaining a continuous degree of uncertainty. In contrast, the smaller architectures frequently expressed absolute certainty that a respondent would abstain. Specifically, the GPT-4.1-nano model assigned exactly $P(\text{voted}) = 0$ to $61.37\%$ of the simulated respondents. 

This divergence in turnout prediction directly cascaded into the secondary task. The smaller models frequently returned undefined party preferences, effectively assigning a probability of zero to all political subjects. As demonstrated in the overlap column of Table~\ref{tab:output_characteristics}, this failure to generate a preference is fundamentally linked to the absolute turnout prediction. The models functionally link the two outputs, actively declining to generate party preferences for respondents they definitively predict will abstain from the election.

\subsection{Aggregate Election Outcomes}

Across all evaluated models, simulated vote shares closely match official election results. The best-performing models achieve MAE values around 2.5 percentage points, comparable to high-quality pre-election polls.

Smaller or lower-capacity models tend to regress toward population averages, while larger models better differentiate sociodemographic profiles.

    \begin{figure}
        \centering
        \begin{minipage}{.48\textwidth}
            \includegraphics[width=0.9\linewidth]{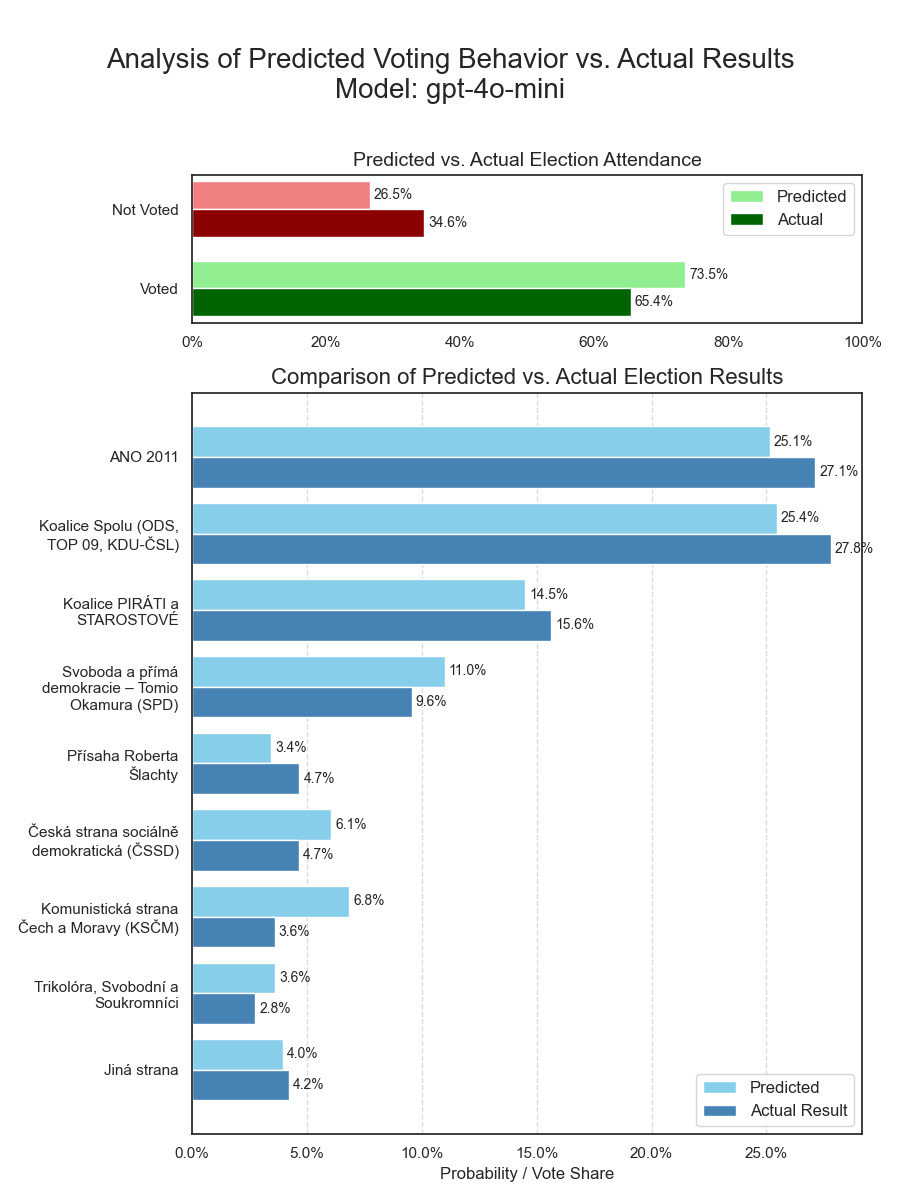}
        \end{minipage}\hfill 
        \begin{minipage}{.48\textwidth}
            \includegraphics[width=0.9\linewidth]{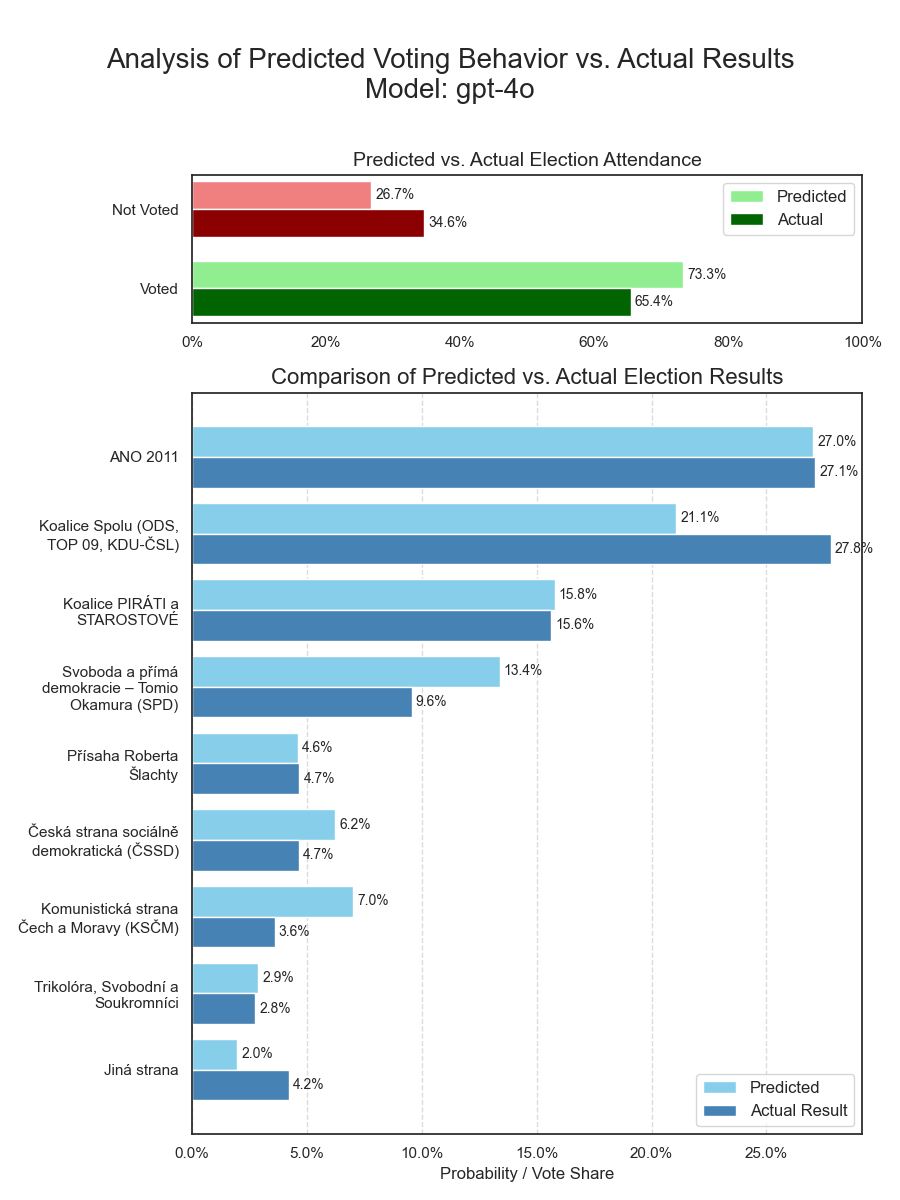}
        \end{minipage}
        
        \caption{Voting simulation results using soft voting, for the GPT-4o-mini model (left) and the GPT-4o model (right).}
        \label{fig:comparison_gpt4o}
    \end{figure}

    \begin{figure}
        \centering
        \begin{minipage}{.48\textwidth}
            \includegraphics[width=0.9\linewidth]{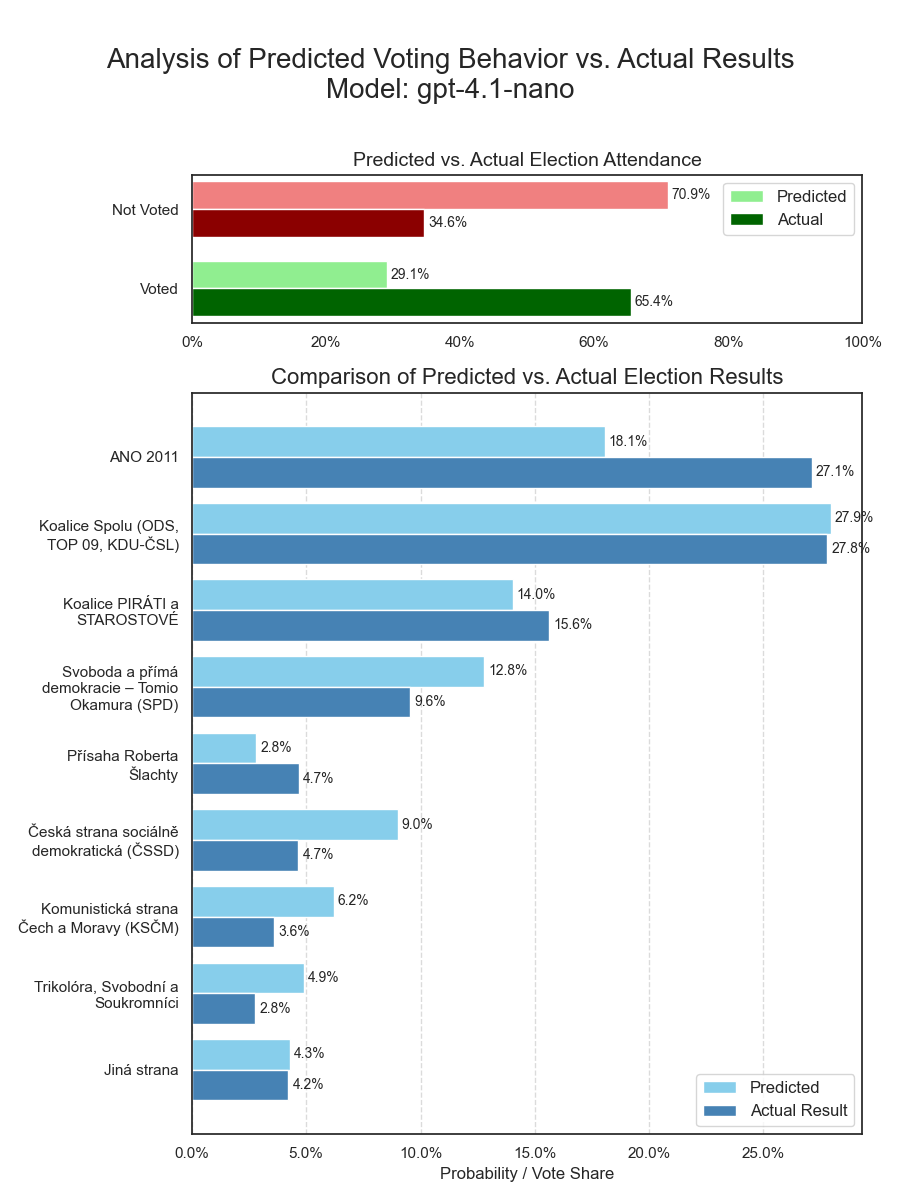}
        \end{minipage}\hfill 
        \begin{minipage}{.48\textwidth}
            \includegraphics[width=0.9\linewidth]{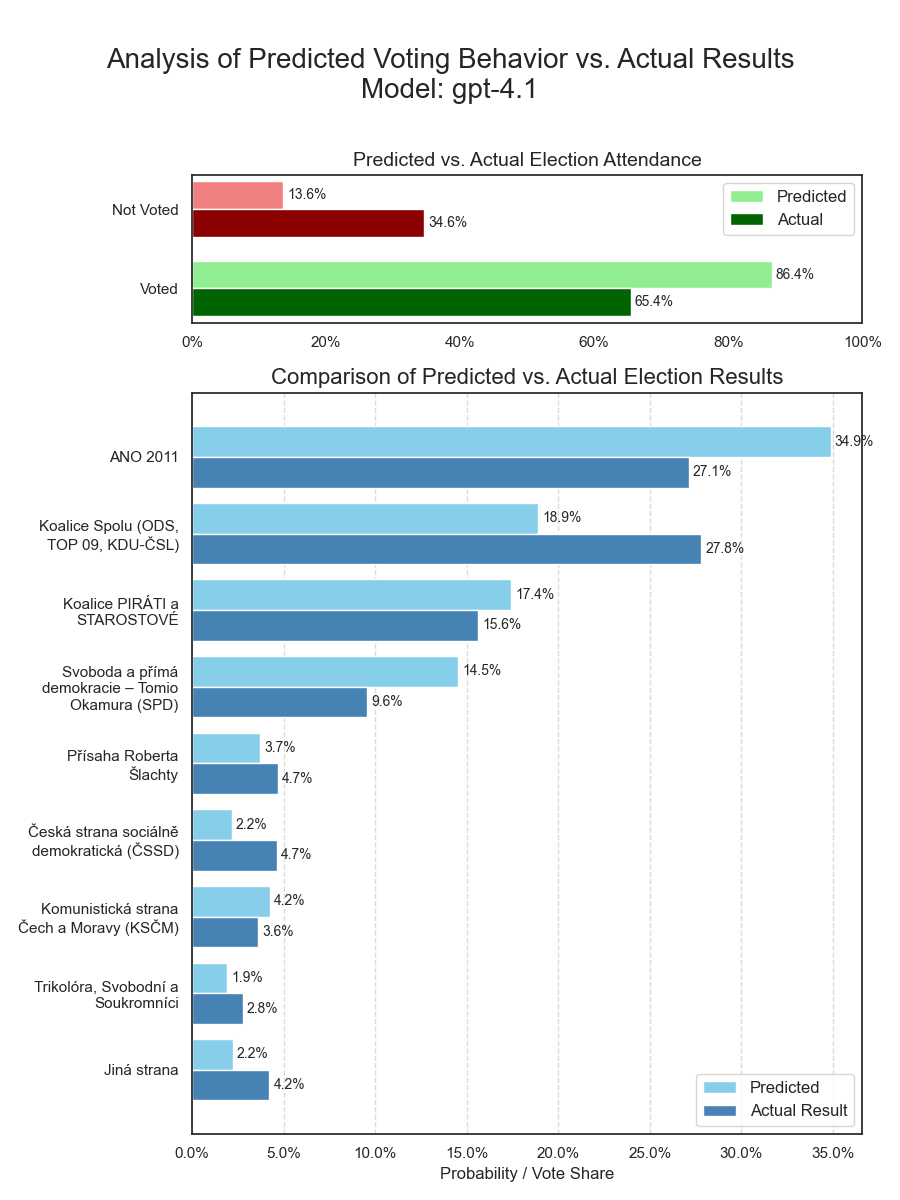}
        \end{minipage}
        
        \caption{Voting simulation results using soft voting, for the GPT-4.1-nano model (left) and the GPT-4.1 model (right).}
        \label{fig:comparison_gpt4.1}
    \end{figure}

\subsection{Regional Accuracy}

Regional MAE analysis reveals that model accuracy is relatively stable across Czech regions, with slightly higher errors in regions with more volatile turnout patterns. This suggests that demographic conditioning captures broad geographic structure but remains sensitive to turnout uncertainty.

    \begin{figure}
        \centering
        \includegraphics[width=\textwidth]{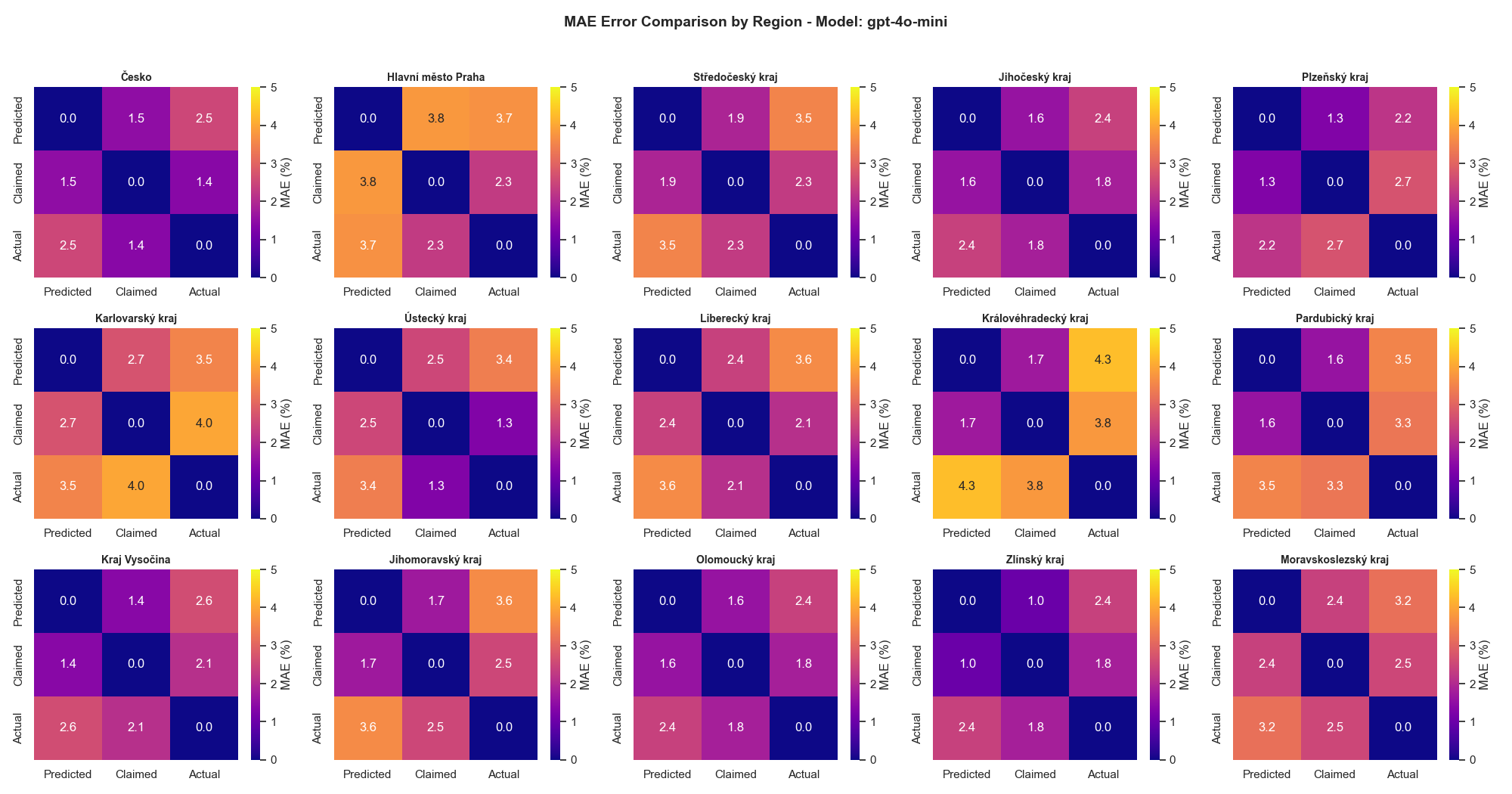} 
        
        \caption{Mean Absolute Error (MAE) between Predicted, Claimed, and Actual voting behaviour in Czechia and across all 14 Czech regions for GPT-4o-mini model.}
        \label{fig:mae_by_region_gpt4omini}
    \end{figure}

    \begin{figure}
        \centering
        \includegraphics[width=\textwidth]{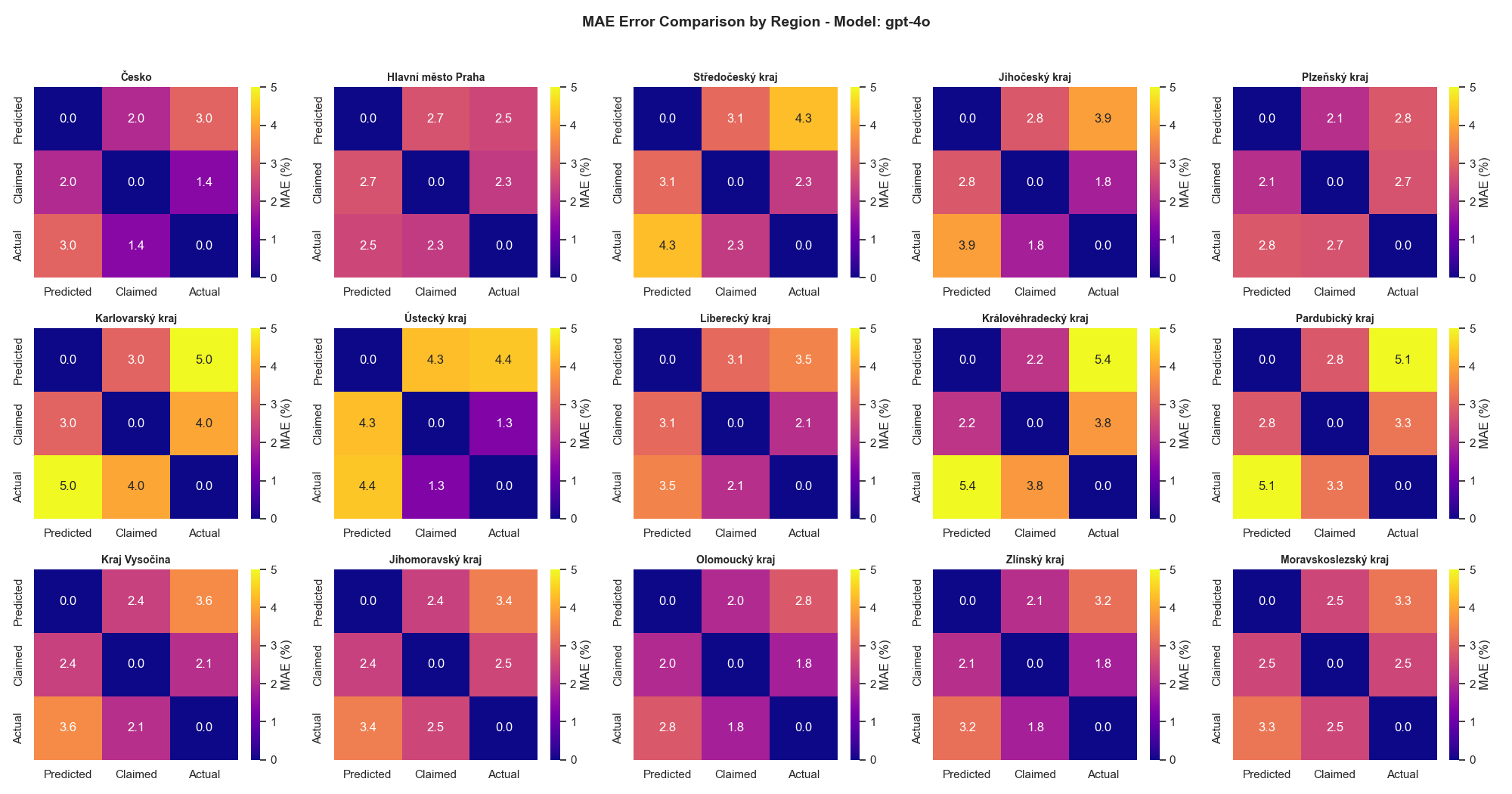} 
        
        \caption{Mean Absolute Error (MAE) between Predicted, Claimed, and Actual voting behaviour in Czechia and across all 14 Czech regions for GPT-4o.}
        \label{fig:mae_by_region_gpt4o}
    \end{figure}

\subsection{Correlation Structure and Political Blocs}

Correlation matrices of predicted vote shares reveal two dominant political blocs:
\begin{itemize}
\item SPOLU and PirSTAN
\item ANO and SPD, with positive association to KSČM
\end{itemize}

These structures closely mirror correlations observed in real municipal-level election data, despite the model having no access to electoral statistics.

            \begin{figure}
                \centering
                \includegraphics[width=0.9\linewidth]{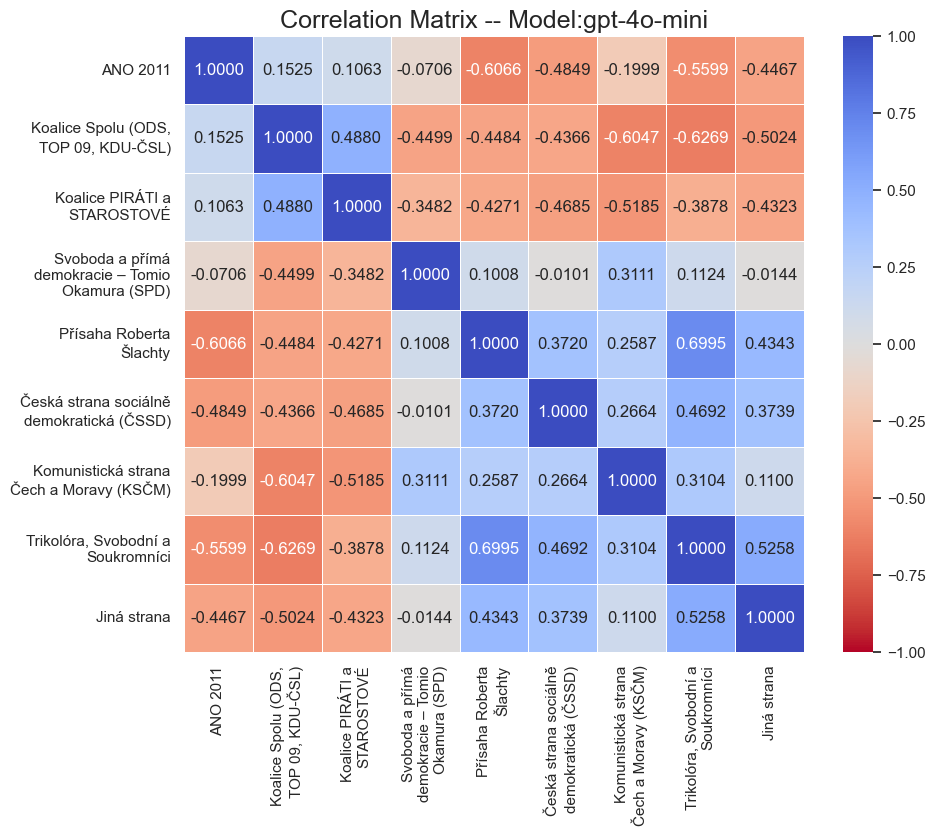}            
                \caption{Correlation matrix of predicted vote shares for the GPT-4o-mini model.}
                \label{fig:corr_matrix_gpt4omini}
    
            \end{figure}
            \begin{figure}
                \centering
                \includegraphics[width=0.9\textwidth]{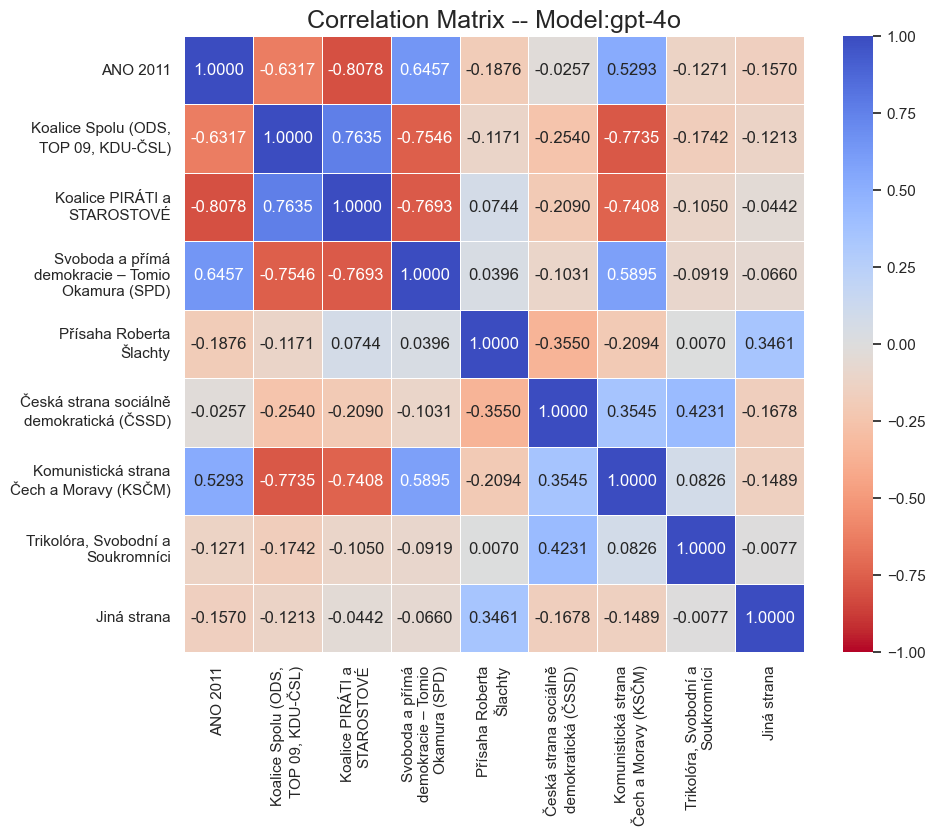}
                \caption{Correlation matrix of predicted vote shares for the GPT-4o model.}
                \label{fig:corr_matrix_gpt4o}
            \end{figure}

\subsection{Structural Validation}

   

To assess whether the simulated outputs recover meaningful structure beyond aggregate accuracy, we conduct several additional validation analyses. Since we treat the outputs of LLMs as probability distributions  we have a linear constraint on the values. Therefore, we perform an isometric log-ratio transformation with respect to the Helmert basis to obtain values which are unconstrained.

\subsubsection{Effect of input variables}

\begin{figure}
    \centering
    \includegraphics[width=0.9\linewidth]{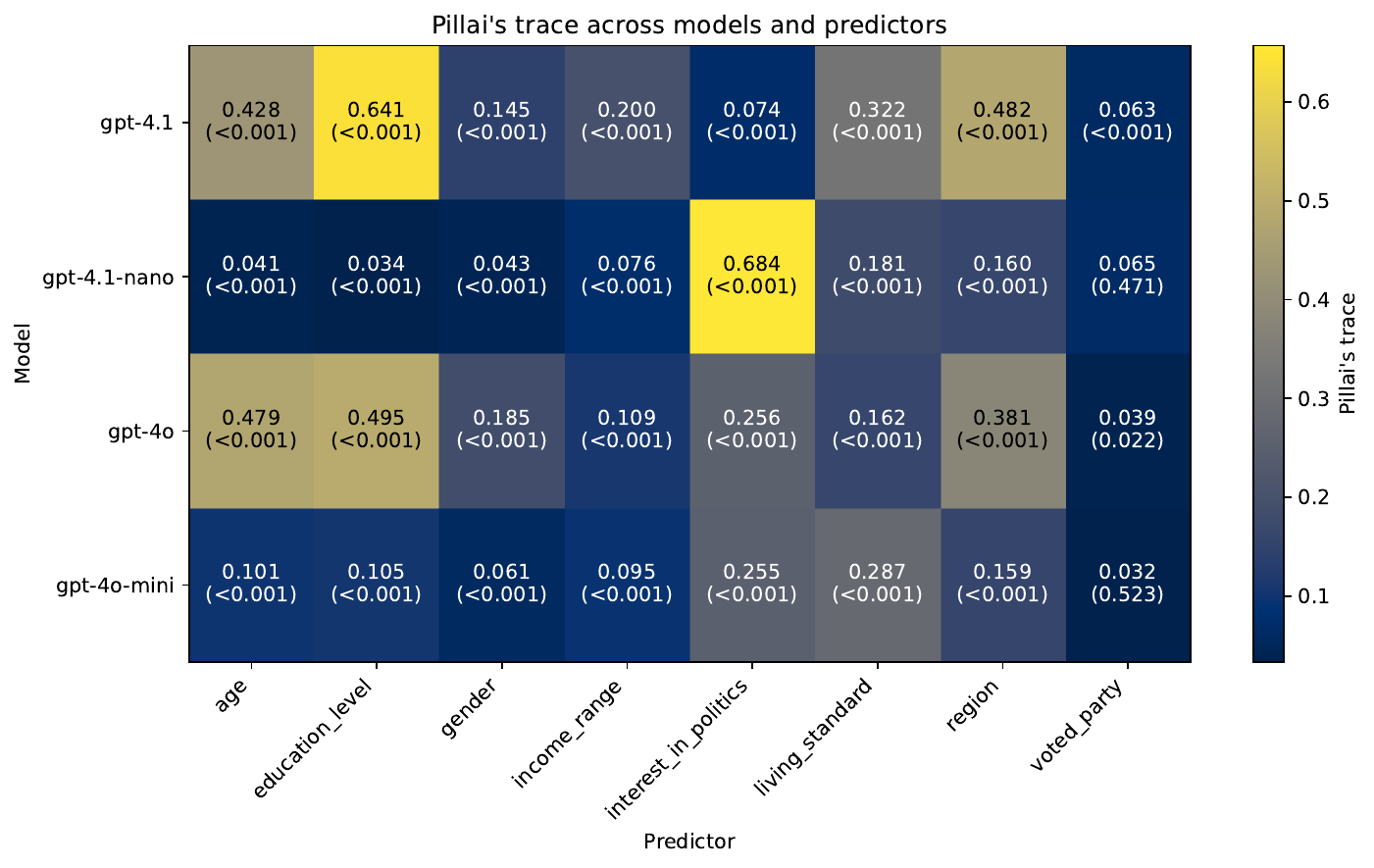}
    \caption{The values are Pillai’s trace statistics, interpreted here as indicators of explained multivariate variance by each variable in MANOVA that was performed for each model. The values in the brackets are the p-values.}
    \label{fig:manova_heatmap}
\end{figure}

To assess if and how much the LLMs actually use the information given in the prompt, we performed for each one the multivariate analysis of variance (MANOVA). The results are in the figure \ref{fig:manova_heatmap}. As a negative control, we also included the ``voted\_party" variable which was {\bf not} used in the prompt. All prompt input variables have a statistically significant small to large effect. The small models have large p-value for the ``voted\_party'' variable which shows either small or heterogeneous effect or no effect whatsoever. This variable is statistically significant for the larger models, but its effect size, as measured by Pillai's trace, is the smallest among the independent variables. 

\subsubsection{Difference to real election outcomes}

To better judge accuracy of LLMs output based on panel data, we compiled a small cross-sectional dataset of Czech parliamentary election voting model estimates from three polling agencies—STEM, Median, and Kantar CZ—fielded in late summer 2021. For each agency, we recorded the reported vote-share (as proportions) for the main electoral lists: ANO 2011; the Spolu coalition (ODS, TOP~09, KDU-\v{C}SL); the PIR\'{A}TI and STAROSTOV\'{E} coalition; SPD; KS\v{C}M; P\v{r}\'{i}saha; \v{C}SSD; Trikol\'{o}ra/Svobodn\'{i}/Soukromn\'{i}ci; and an aggregated residual category (``Jin\'{a} strana'') capturing other parties. Source documents are cited as primary references for each agency estimate \cite{stem2021,median2021,kantar2021}.

\begin{figure}
    \centering
    \includegraphics[width=0.8\linewidth]{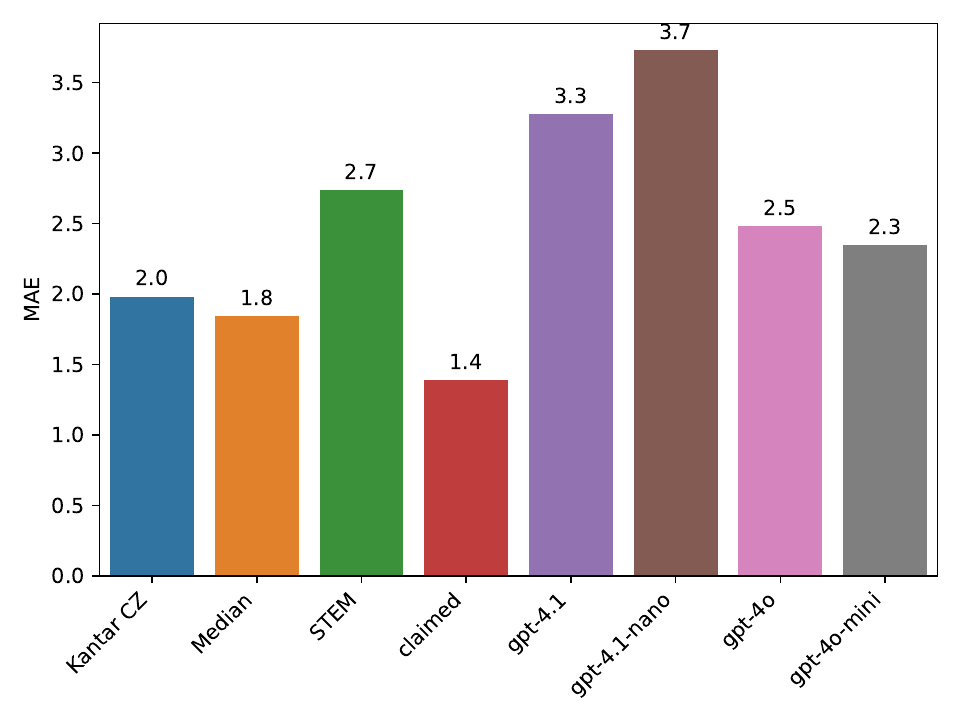}
    \caption{Mean absolute error to the real election outcomes. Columns Kantar CZ, Median and STEM stand for latest pre-election polls of three major Czech polling agencies. The column ``claimed'' is given by the self-reported values from the panel dataset.}
    \label{fig:agentury}
\end{figure}

The mean absolute errors (MAE) for the election outcomes are in figure \ref{fig:agentury}. The figure also contains MAE of the self-reported voted party from our panel data. We see that the results of GPT-4o models are in the range of late pre-election estimates reported by major Czech polling agencies. 

\begin{figure}
    \centering
    \includegraphics[width=0.8\linewidth]{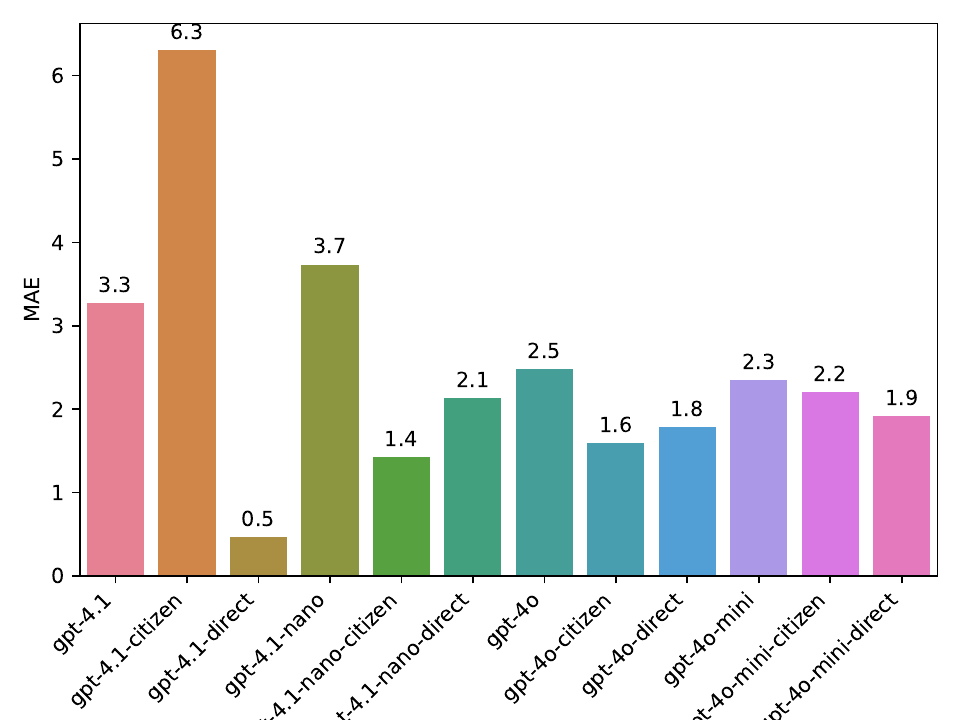}
    \caption{MAE for the latent knowledge which was extracted by two different prompts.}
    \label{fig:latent}
\end{figure}

The mean absolute error of model prior knowledge (as extracted through the ``citizen'' and ``direct'' prompt described in  section \ref{sec:model_prior}) to the election outcomes are depicted in the figure \ref{fig:latent}.

\subsubsection{CHES comparison}

We use party-level data from the Chapel Hill Expert Survey (CHES) 2019 wave \cite{polk2017ches}, which provides expert-coded placements of European political parties on a broad set of ideological and policy dimensions. For the Czech Republic, we rely on continuous issue-position variables capturing economic left--right ideology, social and cultural conservatism--liberalism, European integration, immigration, and related policy domains. These expert assessments are widely used as a benchmark for measuring ideological proximity and party competition. All CHES variables are standardized prior to analysis to ensure comparability across dimensions.

To align CHES data with the Czech party system as it appeared in the 2021 parliamentary election, we aggregated party positions for pre-electoral coalitions. Specifically, we construct composite CHES positions for the \emph{Spolu} coalition (ODS, TOP~09, KDU--ČSL) and for the \emph{Pirates+STAN} coalition using a simple arithmetic mean of member-party positions, assigning equal weight to each coalition partner. This aggregation reflects an assumption of equal ideological contribution by coalition members and avoids introducing external weighting schemes.

\begin{figure}
    \centering
    \includegraphics[width=0.8\linewidth]{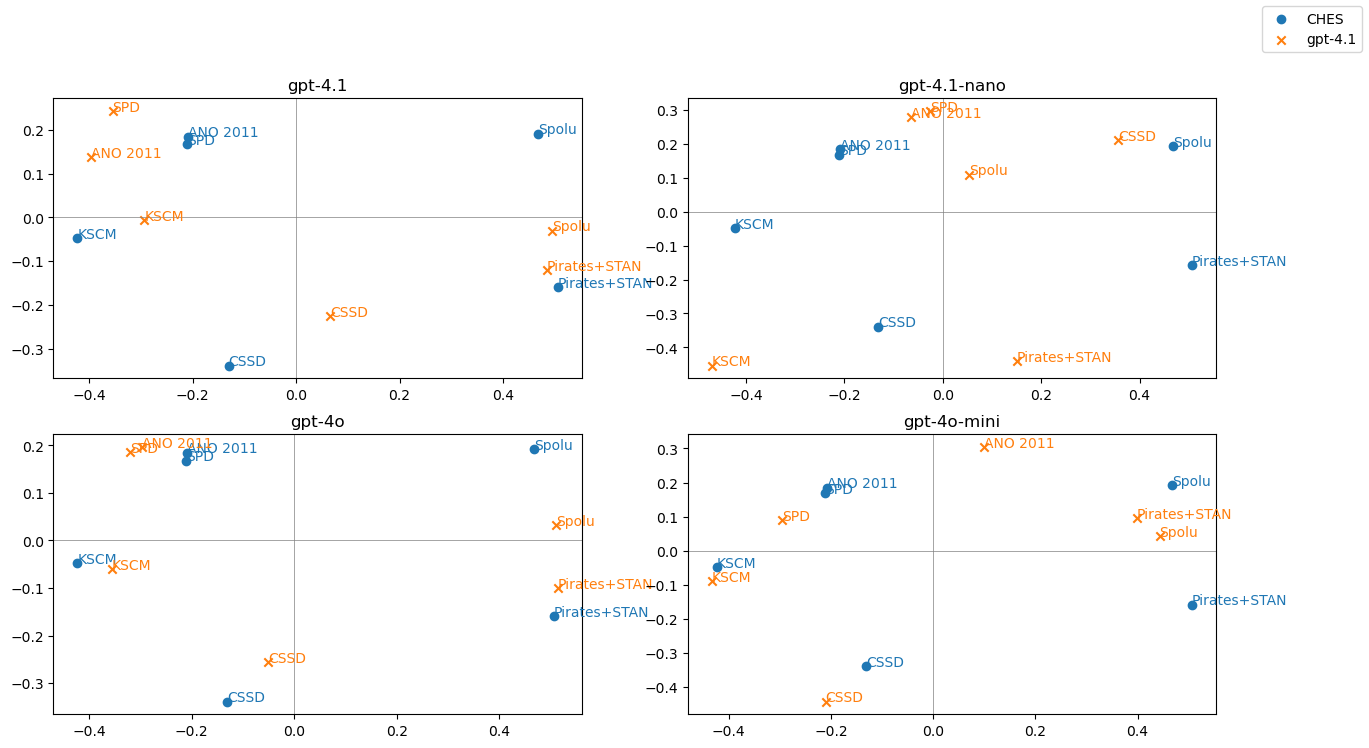}
    \caption{Best 2D alignment between the inter-party distances in the CHES dataset and the LLMs output.}
    \label{fig:procrustes}
\end{figure}

Party--party distance matrices are computed from both the CHES-based positions and the model-implied party correlation structure. Each distance matrix is embedded into a two-dimensional space using classical multidimensional scaling (MDS). To assess structural agreement between  expert-based and model-based geometries, the resulting configurations are aligned via Procrustes analysis, which removes differences in translation, rotation, and scale. The result is shown in  Figure \ref{fig:procrustes}. The best alignment is observed for GPT-4o, where only the Spolu coalition remains visibly misaligned.

\subsubsection{Voter flow validation}

As an independent validation of the model's implied inter-party structure, we compare the correlation matrices of simulated vote shares against empirically documented voter flows from PAQ Research's post-election analysis of the 2021 Czech parliamentary election (\cite{prokop2021paq}). The PAQ study, based on a representative longitudinal panel (N=3,891), tracks individual-level transitions between parties from September 2020 through the October 2021 election, providing detailed diagrams of voter movement.

The PAQ analysis documents three dominant flow patterns. First, a massive transfer of approximately 370,000 voters from PirSTAN to SPOLU in the final campaign phase, reflecting a strategic consolidation of the anti-Babiš electorate around the stronger challenger. Second, ANO absorbed roughly 200,000 voters from ČSSD, KSČM, SPD, and Přísaha in the late campaign, effectively cannibalising its potential coalition partners. Third, PirSTAN and ANO exhibited virtually no direct voter exchange, reflecting their position on opposite sides of the central political divide.

These empirically observed flow patterns correspond closely to the correlation structure recovered by our models without access to any voter-level or flow data. The GPT-4o correlation matrix (Figure \ref{fig:corr_matrix_gpt4o}) shows a strong positive correlation between SPOLU and PirSTAN (0.76), consistent with the large voter transfer between these two coalitions and their shared demographic base of younger, educated, higher-income voters. ANO exhibits strong positive correlations with SPD (0.65) and KSČM (0.53), mirroring the documented absorption of voters from these parties and their overlapping base among older, less educated, and lower-income voters. Conversely, ANO and PirSTAN show a strongly negative correlation (-0.81), consistent with the near-absence of direct voter exchange between these two camps.

The smaller GPT-4o-mini model (Figure \ref{fig:corr_matrix_gpt4omini}) recovers the same qualitative structure, positive SPOLU–PirSTAN association, a distinct ANO–SPD–KSČM cluster, and negative cross-bloc correlations, though with weaker magnitudes, reflecting its generally less differentiated demographic conditioning.
That the models recover voter flow structure from demographic profiles alone suggests that the sociodemographic composition of party electorates is sufficiently distinct to imply the observed inter-party competition patterns. Parties that share demographic constituencies exhibit positively correlated simulated vote shares; parties that draw from opposing demographic pools exhibit negative correlations. The LLMs, in effect, reconstruct the competitive geometry of the Czech party system as an emergent property of demographic conditioning.

\section{Discussion}

\subsection{Why Does This Work?}

The results presented in this paper can be understood through three complementary layers of explanation.

The first is data coverage. Elections are among the most extensively discussed topics in online discourse. Every Czech parliamentary election generates thousands of articles, commentaries, and analyses that explicitly connect demographic characteristics to voting patterns. The training corpora of large language models inevitably contain substantial amounts of such material. In this sense, the models did not need to infer the relationship between demographics and vote choice, they needed only to compress it.

However, compression alone does not explain the results. Our MANOVA analysis (Figure \ref{fig:manova_heatmap}) demonstrates that the models produce systematically different probability distributions for different demographic profiles, education and age consistently show the largest effects across all four models, which aligns with the education-based cleavage documented by Gethin, Martínez-Toledano, and Piketty \cite{gethin2022brahminleft}. More strikingly, the correlation matrices of predicted vote shares (Figures \ref{fig:corr_matrix_gpt4omini} and \ref{fig:corr_matrix_gpt4o}	) recover the two dominant blocs of Czech politics, SPOLU and PirSTAN on one side, ANO and SPD with positive association to KSČM on the other, without any access to electoral statistics. The Procrustes alignment with CHES expert-coded positions (Figure \ref{fig:procrustes}) further confirms that the models' implicit ideological geometry corresponds to independently established party placements. This suggests that the compression preserves internal structure: not merely the marginal distribution of vote shares, but the relational patterns between demographics and parties, and between parties themselves. These findings are consistent with recent mechanistic evidence that political perspective is encoded as approximately linear geometry in model activation space \cite{kim2025linearpolitics}, extending this observation to a non-Anglophone multiparty context.

The third layer is aggregation. Individual-level predictions are noisy, benchmarks suggest that LLMs achieve only 30–40\% individual accuracy in sociological survey simulation (\cite{wang2025sociobench}). The soft voting procedure compensates for this by averaging probability distributions across thousands of profiles, canceling random error and revealing stable structural patterns. The method does not require individual accuracy; it requires that individual errors be sufficiently uncorrelated across party categories.

\subsection{Structure over Accuracy}

It is important to distinguish what the demographic conditioning adds beyond the models' prior knowledge of election results. As Figure 9 shows, all models possess substantial prior knowledge of the 2021 Czech election, GPT-4.1 achieves MAE of 0.5 when asked directly for the result, effectively retrieving the outcome from its training data. The demographic simulation does not always improve upon this aggregate accuracy at the national level.

The value of demographic conditioning lies elsewhere: it adds structure. A direct query yields a single national-level distribution. The demographic simulation yields differentiated predictions across 3,880 individual profiles, enabling regional decomposition (Figures 3 and 4), inter-party correlation analysis (Figures 5 and 6), and diagnostic assessment of which sociodemographic variables drive model behaviour (Figure 7). In this sense, the framework functions less as a prediction tool and more as an instrument for exploring what sociological regularities the model has encoded and how it maps demographic profiles to political preferences.

This reframing also opens the possibility of systematic exploration of the profile space. Because the model accepts any well-formed demographic description, it can be queried with profiles that do not appear in the survey data, including controlled contrasts where a single variable is manipulated while others are held constant. For instance, one can construct two identical profiles differing only in interest in politics, or education level, and examine how the predicted vote distribution shifts. This resembles experimental logic more than survey logic: the "treatment" is a change in the prompt, and the "outcome" is the shift in predicted probabilities. Such counterfactual exploration is confounded in observational data, where interest in politics correlates with education, age, and income. In the model, these variables can be independently manipulated.

The outputs of such explorations reflect the model's encoded associations, not causal reality. As a tool for hypothesis generation, however, the framework has clear practical value: it can suggest that interest in politics shifts the distribution in a particular direction, which can then be tested against empirical data. It transforms the LLM from a black-box predictor into an interrogable map of demographic-to-political associations.

\subsection{Epistemic Status}

The models evaluated in this study do not ``know" why people vote as they do. They possess no causal understanding of political behaviour and no access to individual decision-making processes. Their outputs should be interpreted as simulations of socially plausible behaviour conditioned on demographic descriptions, not as explanations or forecasts.

Epstein \cite{epstein2008whymodel} identifies sixteen reasons to model beyond prediction, including illuminating core dynamics, suggesting analogies, discovering new questions, and disciplining intuitions. Our framework falls squarely within this exploratory tradition. The purpose is not to predict the next election or to replace survey research, but to systematically interrogate what sociopolitical regularities a given model has absorbed from its training corpus and whether these regularities correspond to documented social structures.

Buttrick \cite{buttrick2024compression} proposes viewing LLMs as ``compression algorithms for human culture", systems whose internal representations encode statistical regularities from collective textual output and are themselves worthy of study. This framing captures the epistemic status of our approach precisely. The model's output is a reflection of how the relationship between Czech demographics and voting behaviour is represented in the digital discourse that constitutes its training data. When this reflection aligns with documented electoral patterns, it tells us something about the structure of that discourse; when it diverges, it reveals the limits or distortions of the model's encoding.

\subsection{Limitations}

Several limitations must be acknowledged when interpreting these results.

The method assumes that sociodemographic attributes can be faithfully translated into natural language prompts without introducing unintended framing effects. While care was taken to use neutral, declarative language and to conduct all simulations in Czech, ensuring cultural and linguistic alignment with the electoral context, prompt phrasing may still activate stylistic or cultural associations not present in the original survey variables. The choice to prompt in Czech is both a strength and a potential source of fragility: it ensures appropriate cultural framing, but Czech constitutes a small fraction of typical training corpora, and the quality of the model's Czech-language encoding may differ from its English-language capabilities.

The soft voting aggregation relies on the assumption that individual-level errors are largely uncorrelated and cancel through averaging. If errors are systematic, for example, due to consistent overrepresentation of certain political narratives in training data, aggregate estimates may be biased in ways that are not detectable from aggregate accuracy alone. Bisbee et al. \cite{bisbee2024syntheticreplacements} demonstrate precisely this risk: synthetic survey means can match population averages while regression coefficients diverge significantly, with signs flipping in one-third of cases. Our multi-level validation strategy, comparing not only aggregate vote shares but also covariance structure, CHES-based ideological geometry, and regional breakdowns, partially addresses this concern by testing structural correspondence rather than relying on mean accuracy alone. Nevertheless, the possibility of systematic bias that survives aggregation cannot be ruled out.

The models evaluated in this study align most closely with the perspectives of liberal, high-income, well-educated respondents \cite{santurkar2023whoseopinions}, a bias that may systematically distort the simulation of older, less educated, or more conservative voters. The MANOVA results suggest that smaller models in particular respond less strongly to demographic conditioning, which may reflect a tendency to fall back on a default, and potentially skewed, prior rather than engaging with the specific profile.

The framework was tested on a single election, in a single country, using models from a single provider. The 2021 Czech parliamentary election represents a relatively stable multiparty system with well-documented sociodemographic voting patterns. Results may not generalise to political systems with weaker demographic structuring, higher volatility, or different media ecosystems. Temporal stability is also untested: because the models' training data has a fixed cutoff, the framework captures the state of public discourse up to that point and cannot account for late campaign events, short-term scandals, or rapid opinion shifts. Replication across providers (Claude, Llama, Gemini), elections, and national contexts is necessary before broader conclusions can be drawn.

Finally, an ethical concern must be stated directly. Any system that maps demographic profiles to political preferences is, in principle, repurposable for political microtargeting. The framework presented here operates at the aggregate level and is intended as an exploratory research tool, but the underlying capability, conditioning on a demographic description to obtain a predicted vote distribution, could be misused for individualised persuasion or manipulation. We believe that transparency about this capability is preferable to obscurity, and that the appropriate response is not to avoid the research but to clearly delimit its intended scope and to advocate for responsible use.

\subsection{Elections as Easy Case: Future Directions}

Elections represent a uniquely favourable test case for LLM-based sociological simulation. Electoral behaviour is massively represented in online discourse, the outcome variable is categorical and well-defined, and the relationship between demographics and vote choice is explicitly discussed in thousands of media articles, academic publications, and public commentaries. The models' training corpora are, in effect, saturated with this particular mapping. That the framework works for elections does not mean it will work for less salient sociological phenomena.

The genuine test of LLMs as implicit sociological models lies in domains where the textual footprint is smaller and the relationships more subtle. Do models encode the distribution of personality traits across demographic groups? Can they reconstruct patterns of institutional trust by region and education? Do they capture the sociodemographic gradients of susceptibility to conspiracy narratives? These questions involve associations that are documented in the academic literature but far less prominent in everyday online discourse. The survey underlying this study, "Society of Distrust," with over 130 questions spanning trust, media consumption, and conspiracy beliefs, offers a direct path toward such testing. If LLMs have absorbed sociological regularities in general, they should reconstruct these distributions with reasonable fidelity; if they have absorbed only the most media-salient political narratives, they will fail.

Further directions include cross-provider comparison to assess whether the encoded regularities are model-specific or shared across architectures, temporal analysis using the same prompts on successive model versions to measure drift in encoded associations, and systematic sensitivity analysis of prompt design, including permutations of variable ordering, phrasing, and language choice. The framework proposed here is deliberately general: any domain where a structured profile can be mapped to a probabilistic outcome is, in principle, amenable to the same approach.

\section{Conclusions}

This paper introduced and empirically evaluated a methodological framework for using large language models as implicit sociological models of aggregate political behaviour. By conditioning LLMs on individual-level sociodemographic profiles and aggregating probabilistic outputs through a soft voting procedure, we demonstrated that contemporary models can reconstruct key features of real-world electoral outcomes without access to electoral statistics.

Using the 2021 Czech parliamentary election as a validation case, the framework reproduced official vote shares with low aggregate error, recovered known political bloc structures, and aligned with independently established sociodemographic voting gradients. Importantly, these results emerge not from explicit political knowledge or causal reasoning, but from statistical regularities encoded in large-scale textual data.

The contribution of this work is methodological rather than predictive. The proposed approach does not explain why individuals vote as they do, nor does it offer real-time forecasting or individual-level inference. Instead, it provides a systematic way to interrogate the latent social knowledge embedded in language models and to assess its correspondence with observed social structures.

Within appropriate epistemic and ethical boundaries, LLM-based demographic simulation may serve as a complementary exploratory tool for computational social science. Potential applications include hypothesis generation, robustness checks for survey-based findings, and comparative analyses across models or contexts. Future work should focus on replication across countries and electoral systems, sensitivity analyses of prompt design, and deeper investigation into the sources and limits of sociopolitical regularities encoded in language models.

\section*{Declarations}

\subsection*{Data and code availability}
Data and code are available at \url{https://github.com/11NIKTO11/Panel\_LLM}

\subsection*{Acknowledgments}
This work was supported by the European Regional Development Fund project ``Beyond Security: Role of Conflict in Resilience-Building" (reg. no.: CZ.02.01.01/00/22\_008/0004595). 
Data for this research was supported by the NPO ``Systemic Risk Institute" no. LX22NPO5101, funded by European Union – Next Generation EU (Ministry of Education, Youth and Sports, NPO: EXCELES).

\subsection*{Conflict of interest}
The authors declare no conflict of interest.

\bibliography{references-final}

\begin{appendices}

\section{Prompt Template}

Each respondent was converted into a natural-language prompt encoding sociodemographic attributes. The general template was as follows:

\begin{quote}
I am a \{gender\}, \{age\} years old. My highest completed education is \{education\}.  
I live in the \{region\} region, \{district\} district, in a municipality with \{municipality size\} inhabitants.  
My employment status is \{employment\}. Our household income is \{income range\}.  
I consider our living standard to be \{living standard\}.  
I am \{interest in politics\} interested in politics.  

In the 2021 parliamentary elections, I voted for:
\end{quote}

All categorical values were rendered in plain, neutral language to minimize stylistic cues.

\section{Output Schema}

Models were constrained to return structured outputs consisting of:

\begin{itemize}
\item Probability of turnout:
\[
P(\text{voted}), \; P(\text{not voted})
\]
\item Conditional probability distribution over party choices:
\[
P(p \mid \text{voted}), \quad p \in \{\text{ANO}, \text{SPOLU}, \text{PirSTAN}, \text{SPD}, \text{Přísaha}, \text{ČSSD}, \text{KSČM}, \text{Trikolóra}, \text{Other}\}
\]
\end{itemize}

Probabilities were required to sum to one within each component.

\section{Soft Voting Aggregation}

Let $i$ index respondents and $p$ political parties.  
Each respondent contributes a weighted vote:

\[
w_{i,p} = P(p \mid i) \cdot P(\text{voted} \mid i)
\]

Aggregate party support is computed as:

\[
\hat{V}_p = \frac{\sum_i w_{i,p}}{\sum_{p'} \sum_i w_{i,p'}}
\]

This approach preserves uncertainty at the individual level while producing stable aggregate estimates.

\section{Ethical and Methodological Scope}

This methodology is not intended for individual-level inference, prediction, or persuasion.  
Simulated outputs reflect statistical regularities in training data and may reproduce historical biases.  
All analyses are conducted at the aggregate level and evaluated against publicly available benchmarks.

The method should be understood as exploratory and descriptive, complementing—not replacing—traditional survey-based research.

\end{appendices}

\end{document}